\documentclass{JHEP3}
\usepackage{amsmath,cite}
\usepackage{epsfig}
\input epsf
\catcode `@=11 \@addtoreset{equation}{section} \catcode `@=12
\def\be{\begin{equation}}
\def\ee{\end{equation}}
\def\ba{\begin{array}{l}}
\def\ea{\end{array}}
\def\bea{\begin{eqnarray}}
\def\eea{\end{eqnarray}}

\def\del{\partial}

\def\sl2{$SL(2,R)$}

\title{\bf D-branes in 2D Lorentzian Black Hole}

\author{{\sc K. P. Yogendran \footnote{patta@mri.ernet.in, 
patta\_yogendran@yahoo.co.in}}\\
Harish-Chandra Research Institute\\
Chhatnag Road, Jhusi\\
Allahabad, INDIA 211 019
}

\abstract{ We study D-branes in the Lorentzian signature 2D black hole
in string theory. We use the technique of gauged WZW models to
construct the associated boundary conformal field theories. The main
focus of this work is to discuss the (semi-classical) world-volume
geometries of the D-branes. We discuss comparison of our work
with results in related gauged WZW models. }

\preprint{HRI-P-04-06-002\\hep-th}
\begin{document}

\section{Introduction}
In view of the recent activity in two-dimensional string theory much
of it revolving around the new interpretation of the $c=1$ matrix
model as (a scaled limit) of the open string theory on unstable
D0-branes in Liouville theory, it is natural to study the possible
branes in other 2-dimensional string backgrounds. The case of the
Euclidean black hole has been discussed in a recent paper
\cite{SR}. In this work, we investigate the Lorentzian black hole
string theory using primarily semi-classical methods based on the
$SL(2,R)/U(1)$ conformal field theory. A more precise study will
entail the construction of the corresponding boundary states
\cite{KPY3}.

Another motivation in starting this study was that
our prelimnary investigation revealed that the D-branes in the
Lorentzian geometry are time-dependent. Since time dependent solutions
to string theory are somewhat at a premium, such D-branes are likely
to be interesting.

In view of conjectures that the Lorentzian black hole geometry is
related to the phase space picture of the $c=1$ matrix model
\cite{Sumit,Dhar}, such branes could help in studying these questions.

D-branes have proved to be extremely useful in string theory in
general in studying the geometry of backgrounds, especially near
singularities. It is but natural to attempt to study black-hole
singularities using D-branes. Our work could be regarded as a first
step towards such an end.

The outline of this paper is as follows. In the first section, we
provide a short summary of the construction of this string theory
background as the coset $SL(2,R)/U(1)$. This coset construction allows
us to use the the geometrical methods of
\cite{Gawedzki,Elitzur,Sarkissian,KPY2,Walton} to analyse the allowed
boundary conditions consistent with conformal invariance. This general
procedure is described (albeit rather briefly) in the second section.

The third section describes the geometry of the various branes found
by the preceding technique in this string background. It is to be
noted that the semi-classical geometrical analysis presented in this
section is valid only for large level of the CFT. The exact string
background is obtained by setting the level $k=\frac{9}{4}$, when loop
corrections are substantial \cite{DVV, Tseytlin}. The idea then, is to
use the semiclassical understanding to construct the boundary states
along the lines of \cite{Fuchs, MMS,SR}. We describe the various
allowed boundary conditions, and describe the world-volume geometries
of the of the branes. We find three kinds of D-branes namely, D(-1),
D0- and D1-branes which are both emitted from the white hole and fall
into the black hole. We also present a mini-superspace analysis of the
spectrum of excitations of these branes, and attempt to analyse their
stability.

In the next sections, we compare these D-branes with those found in
the Euclidean black hole geometry and that of the Parafermion
theory. In the latter case, one is really comparing boundary CFT's
(not string backgrounds), but we deem this a worthwhile exercise since
this could be useful in constructing the boundary states for these
branes.

Recalling that the extended (Lorentzian) black hole geometry possesses
a duality (a T-duality) \cite{DVV}, which exchanges the region in
front of the horizon with the region behind the singularity, we then
briefly examine the duality relations between these branes.

Lastly we include a summary and list some natural questions and
directions for further study.

The appendix contains some useful co-ordinate charts for $SL(2,R)$
and a brief discussion of the BCFT lagrangian for gauged WZW models.

\section{D-branes in a gauged WZW model\label{ONE}}

D-branes in gauged $G/H$ WZW models have been studied in a series of
papers including \cite{MMS, Fredenhagen, Ishikawa, Gawedzki,
Elitzur,G2, Walton}. The conclusions may be summed up by the
statement: the allowed Dirichlet boundary conditions for open strings
consist of products of (twined) conjugacy classes of $G$ with those of
$H$ projected down to the coset (keeping track of the product factors
and after possible translations by elements of $H$).

This result is explained \cite{Gawedzki,Elitzur,Walton,KPY2} as follows. For
the present purpose, we shall assume that $H$ is the subgroup $H\times
H^{-1}$ or $H\times H$ of the $G\times G$ symmetry of the WZW model -
the former is the vectorial gauging and the latter the axial gauging
of $H$. We will also assume that $H$ is abelian (the case relevant for
the black-hole is $H=U(1)$), in which case the conjugacy classes of $H$ are
the various points of $H$.

The boundary conditions consistent with the symmetries of the ungauged
WZW model are those for which the worldsheet boundaries are restricted
to the (twined) conjugacy classes of the group manifold $G$
\cite{Fuchs}. In this case, the regular conjugacy classes and the
twined conjugacy classes correspond to A-type and B-type branes.

When this sigma model is gauged the allowed boundary conditions must
be consistent with the symmetry being gauged. This can happen in the
following way. For specificity, we shall consider the axial
gauging. In this case, consider the twined conjugacy class $C^{\omega}
_g = \{ \omega(h)\, g\, h^{-1}, \forall h\in G\}$ where $\omega$ is an
outer automorphism that acts on $H$ as $h\in H \rightarrow
\omega(h)\equiv \omega\,h\,\omega^{-1}=h^{-1}$ i.e $\omega$ takes $h$
to its inverse.  Under the axial symmetry
\[C^{\omega} _g\rightarrow k\,C^{\omega} _g \,k=\{\omega(k^{-1} h)\, g\,
(h^{-1}\, k) \,|\forall h\in G\}\] and hence this set of boundary
conditions is left invariant. We could have also translated the
$C^{\omega} _g$ by elements $k_0\in H$ as $k_0 C^{\omega} _g$ (since
$H$ is assumed to be abelian; in the general case one considers products of
conjugacy classes of $G$ and $H$). 

The gauging operation results in a target space which is a coset under
the equivalence relation $g\sim h\,g\,h$, and hence the set of
boundary conditions becomes the projection of $k_0\,C^{\omega} _g$ to
the coset. Note that there is no loss of generality in restricting to
left translations (right translations are equivalent to left
translations). 

This set of boundary conditions does not preserve the $H\times H^{-1}$
current algebra symmetry (target space isometry) of the gauged sigma
model (the twisted conjugacy class is not invariant under $C^{\omega}
_g\rightarrow k\,C^{\omega} _g \,k^{-1}$). Since this symmetry is
spontaneously broken there are zero modes corresponding to
translations along the isometry direction (this is what corresponds to
the left translations by elements $k_0\in H$). We will call these branes
A-type (analogous to the A-branes of \cite{MMS} because the A-branes
of that paper are of A-type w.r.t the vectorial gauging).

On the other hand, we could also consider the regular conjugacy
classes of G. Under the action of the axial symmetry, the conjugacy
classes are left translated by elements of $H:\,$
$l_0\,C_g\rightarrow k\,l_0\,\,C_g \,k=kl_0 kC_g=C^{\omega,H}_{l_0} C_g$
(for generality we have included a translation by an element $l_0\in
H$; the translated set is then a product of a twined conjugacy class
of $H$ and the regular conjugacy class of $G$)
Thus, if we consider the set
\[ \tilde C_g= \cup_{k} \{k\, l_0\,C_g\,k\,|\, k\in H\} \]
this set of boundary conditions is invariant under the axial
gauging. As before, we can translate by elements $k_0\in H$ (and again
left and right translations are equivalent). Observe
that the union is over a fixed G-conjugacy class (which is assumed to
be connected). This set is then projected down to the coset by the
gauging. In this projection, the final brane world-volume that emerges
is the restriction of the regular conjugacy class $C_g$ to the coset.
These B-type brane-worldvolumes are invariant under the isometry.

We can also understand the presence of the two types of branes by
looking at quantum states in the parent theory. The A-type branes are
the D-brane states of the parent theory which are invariant under the
symmetry being gauged. Geometrically this simply means that the
A-brane world-volumes are preserved under the symmetry being gauged.
The other set of branes of the coset theory are obtained simply by
superposing branes of the parent theory (that are {\em not} invariant
under the gauge symmetry) to construct states which are invariant
under the gauge symmetry.

On the other hand, if we consider the vectorial gauging of $H$, then
the roles of the A-type and the B-type branes are reversed. The
regular conjugacy classes are left invariant by the gauge symmetry,
while the twined conjugacy classes are translated. Thus by similar
arguments, we can construct branes in the coset theory, as projections
of (twined) conjugacy classes after suitable translations (and
superpositions).

When the symmetry that is being gauged is the axial symmetry, the two
types of branes preserve different amounts of the target space
isometry of the coset (which is the vectorial symmetry but possibly
anomalous).  The A-type branes arise from the twined conjugacy classes
-- which break the vectorial symmetry in the parent theory
itself. Hence in the coset, they are not invariant under the
isometry. But because they break this symmetry spontaneously, there is
a family of such states 

The B-type brane world-volumes are invariant
under the vectorial symmetry in the parent theory, and hence in the
coset theory, their world-volume is invariant under this symmetry.

We can write down the BCFT sigma model lagrangian explicitly and show
that these boundary conditions are consistent with conformal
invariance (for a brief discussion, see the Appendix). Note however
that these sets of boundary conditions preserve one half of the
current algebra of the bulk gauged sigma model. This is perhaps
sufficient to establish the claim of conformal invariance. 


\section{The Lorentzian black hole \label{bh}}

The Lorentzian black hole is obtained by gauging a non-compact axial
$U(1)$ symmetry of the $SL(2,R)$ WZW model. We shall briefly outline
the procedure -- for details refer to \cite{Wittenbh,DVV}.

The symmetry that is being gauged corresponds to a hyperbolic subgroup
of \sl2 which acts on $g
= \left(\begin{array}{cc} a & u\\-v&b \end{array} \right) 
\in$ \sl2 as 
$\delta g=\epsilon (\sigma_3\,g+g\,\sigma_3)$, i.e. 
\bea
\delta a &=& 2\epsilon a \,\,\,\, \delta u = 0 \\\nonumber
\delta b &=& -2\epsilon b \,\,\,\, \delta v = 0
\eea

In gauging the \sl2 theory, one has to choose a gauge fixing
condition. As Witten has argued, this is a subtle issue since there is
no single gauge choice which gives rise to a globally two dimensional
target space. 

In the region $(1-uv)>0$, $ab>0$ and hence a natural gauge fixing
condition is $a=b$. Upon integrating out the gauge fields (which
appear quadratically), we obtain the sigma model action
\be
L=-\frac{k}{4\pi}\int d^2x\sqrt{h}\, \frac{h^{ij}\,\del_i u\,\del_j v
}{(1-u\,v)}
\ee

In the region $(1-uv)<0$ however, a good gauge fixing condition is
$a=-b$ (because $ab<0$). When $uv=1$ either $a=0$ or $b=0$ or both,
hence we cannot gauge transform a field configuration to the gauge
slice (for either gauge choice).  Although the gauge fixing condition
is singular, the sigma model is itself non-singular.

The target space geometry of the sigma model so obtained is as shown
in the figure \ref{bhole}

\FIGURE{\begin{picture}(300,150)(0,0)
\put(30,0){\epsfig{file=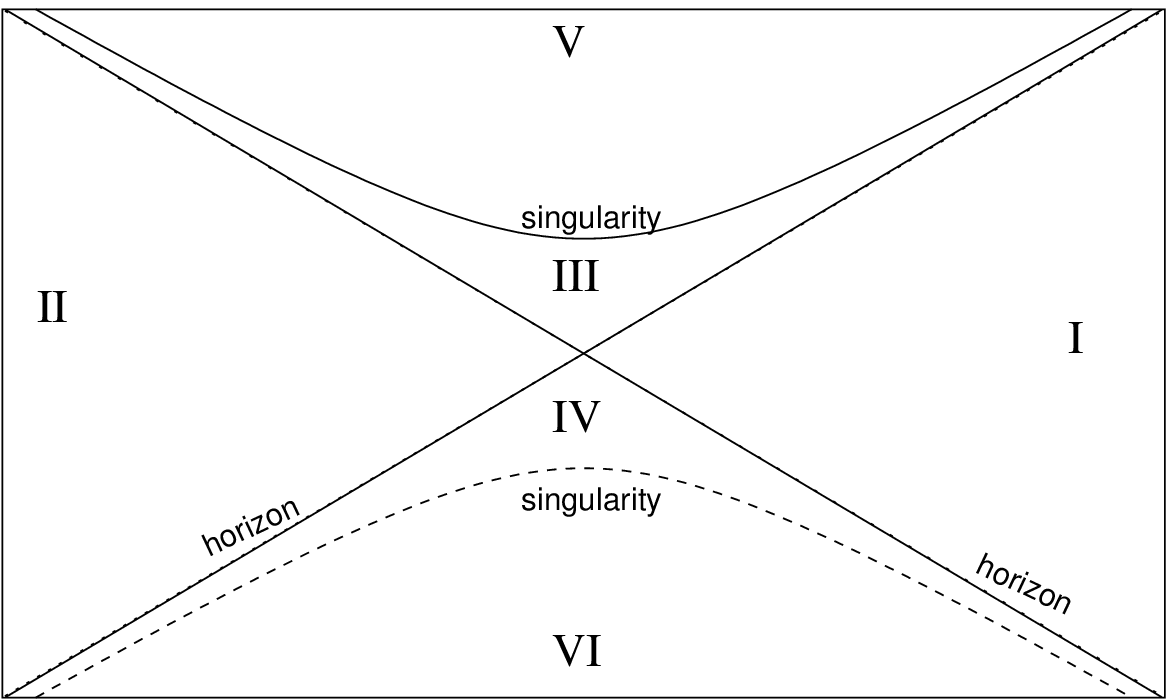,height=5cm}}
\end{picture}
\caption{The Lorentzian black hole geometry.
\label{bhole}}}

In the figure, the diagonal lines $uv=0$ form the horizon, while
$uv=1$ is the singularity (the Ricci scalar diverges as $R\sim
(1-uv)^{-2}$). Regions I and II are asymptotically flat regions and in
regions V and VI time flows ``sideways''. Constant time slices (in
the asymptotically flat region) are straight lines passing through the
origin with time increasing from top to bottom. Thus the black hole
singularity is in the fourth quadrant (in the figure the diagonal
lines are the co-ordinate axes!).

Requiring conformal invariance generates a dilaton at one loop
\be
\Phi=\Phi_0-\frac{1}{2}\ln (1-uv)
\ee
where the parameter $\Phi_0$ is related to the ADM mass of the black
hole. 

We can also proceed slightly differently by gauging the vectorial
action of $H$ \cite{DVV}, i.e which acts on $g\in$ \sl2 as $\delta
g=\epsilon (\sigma_3\,g-g\,\sigma_3)$, i.e. 
\bea
\delta u &=& -2\epsilon u \,\,\,\, \delta a = 0 \\\nonumber
\delta v &=& 2\epsilon v \,\,\,\, \delta b = 0
\eea
Thus a natural gauge choice in this region is $u=v$, and as before we have a
gauge singularity at $ab=1$. In this case, the black hole geometry is
covered by the $(a,b)$ coordinates of \sl2 (the sigma model so
obtained has the same target space metric as before, with $(a,b)$
replacing $(u,v)$ and $ab=1$ being the singularity). 

Thus, we have two descriptions of the black hole geometry: one
obtained by gauging the axial $U(1)$ and another obtained by gauging
the vectorial $U(1)$. These two descriptions are dual to each other
\cite{Giveon, DVV}. We will make use of both descriptions. 

While this analysis is performed at the leading order in $\alpha'$,
the exact background to all orders is known \cite{DVV, Tseytlin,
Teo}. However we shall restrict our investigation to the leading
order. The geometrical description presented in the subsequent
sections makes sense at large $k$. The exact string background is
obtained when $k=\frac{9}{4}$ when $\alpha'$ corrections are
substantial.  It will be very interesting to properly understand what
happens to the D-brane open string CFT in the exact description, in
particular when we reach the singularity and try to continue past it.

In the following, it will be convenient to use another set of
co-ordinates to cover the black hole -- which are natural from the
\sl2 point of view as described in the appendix. In the regions I and
II, the co-ordinate transformation is
$u=-\sinh\rho\,e^{-t},\,\,v=\sinh\rho\,e^{t}$ and the metric is 
\be
ds^2=k(d\rho^2-\tanh^2\rho\,dt^2)
\ee
Analytically continuing $t\rightarrow i\tau$, we get the Euclidean
black hole which also has a description as a gauged WZW model. 

The regions $0<uv<1$ of the black hole can be analytically continued
to the Parafermion (Pf) theory $SU(2)/U(1)$. In this region (III \& IV
in the figure), the co-ordinate change is
$u=\sin\rho\,e^{-t},\,\,v=\sin\rho\,e^{t}$ and the 
metric can be written as 
\be 
ds^2= -k(d\rho^2 -\tan^2 \rho dt^2) 
\ee
The analytic continuation is performed by $t\rightarrow i\tau$ and
$k\rightarrow -k$ which gives us the target space of the parafermion
theory at level $k$.
This target space is topologically the unit disk, and has a
curvature singularity at the boundary of the disk. In this
continuation, the black hole singularity which is a curvature
singularity maps to the boundary of the disk and the horizon of the
black hole maps to the center of the disk.

The ``natural'' co-ordinates for regions V \& VI behind the singularity
are $u=\pm\cosh\rho\,e^{-t},\,\,v=\pm\cosh\rho\,e^{t}$. 

\section{D-branes in the Lorentzian black hole\label{TWO}}

Using the procedure outlined in section \ref{ONE}, and the {\em axially}
gauged WZW construction of the black hole CFT, we can identify the
various D-branes obtainable as BCFT's in this geometry.

The regular conjugacy classes of \sl2 are well known and their
topologies have been described in e.g. \cite{Stanciu}.  \sl2 has only
one outer automorphism upto conjugation \cite{BPet}, which we take to be
conjugation by $\sigma_1$, and hence a one parameter family of twined
conjugacy classes
\[ C^{\sigma,G}_g=\{\sigma_1 h\sigma_1 gh^{-1}, h\in G\}\]
These are characterised by a class invariant $Tr(\sigma_1
g_0)=2\kappa$ and form a connected submanifold in \sl2 for each
$\kappa$. The other outer automorphisms which are \sl2 conjugates of
$\sigma_1$ give rise to twined c.c that are translates of the ones
above. Hence, it is sufficient to restrict our attention to these and
their translates.

Since the twined conjugacy classes are preserved under the gauge
symmetry, their geometry is determined simply by projection to the
coset. These can also be (left) translated by elements in $H$. Thus
they are characterised by two parameters $\kappa$, the class
invariant, and the translation $l_0\in H$ and give rise to A-branes in
these theories. 

The B-type branes are obtained by projecting the set
$C^{\omega,H}_{l_0} C^G _g$ where $C^G _g$ is a regular c.c of
$G=SL(2,R)$. These are characterized by one real parameter, the trace
of $g$ (translations by $l_0$ leave the brane invariant).


We will also have occasion to use the {\em vectorially} gauged WZW
model description. In this case the regular conjugacy classes $C^G _g$
simply project down to the coset, being invariant under the gauge
symmetry. These can also be (left) translated by $l_0\in H$. Thus we
get branes characterised by the trace $Tr(g)$ and by $l_0$. These are
the A-type branes of this theory. 

The B-type branes are obtained by projections of products of $H$ and
$G$ twined conjugacy classes $C^{\omega,H}_{l_0} C^{\omega,G} _g$, and
are characterized by one parameter (the trace $Tr(\sigma_1 g)$; the
$l_0$ translations leave these branes invariant).

As described in the previous section, the region in front of the
horizon can be analytically continued to the Euclidean black
hole. The D-branes in this latter geometry were described by
\cite{SR}, using an $SL(2,C)/SU(2)$ coset description.

Similarly, the region between the horizon and the singularity can be
analytically continued to the Pf-theory. In this case, the D-branes
were investigated in \cite{MMS}.

Thus, we can analytically continue the D-brane world-volumes in these
two regions and compare with the results obtained in these works.  It
is also to be noted that by analytically continuing the Lorentzian
branes, we need not get {\it real} branes in the Euclidean theories
\footnote{We are grateful to Ashoke Sen for pointing this out}.



\subsection{D(-1) branes}

These branes correspond to the identity conjugacy class (c.c). In the
coset, this conjugacy class projects to $(u,v)=(0,0)$.  As described
earlier, left translation by a boost $l_0=diag(e^{t_0},e^{-t_0})\in H$
does nothing to the location of the instanton, and hence we have a
single brane sitting on the (intersection of the future and past)
horizon.

Another way to argue that all these ``boosted'' branes are equivalent
is that in constructing the boundary state for these branes, the
closed string one-point functions are obtained by evaluating the
closed-string primaries at the location of the (point-like)
branes. Since the boost does nothing to location of the branes, the
boundary states are identical, and hence we have just one brane. 

There is another family of such branes, which are seen in the
vectorial gauging. In this case, the identity conjugacy
class maps to the point $(1,1)$. Upon translation by $h_0$, these
branes are translated to $(a,b)=(e^{t_0},e^{-t_0})$ still residing on
the singularity $ab=1$. A different way to assure ourselves of their
existence is by comparison with the parafermion theory as discussed
later in section \ref{comp}.

A similar set of pointlike branes are obtained when one considers the
conjugacy class of $-I\in SL(2,R)$.


\FIGURE{\begin{picture}(300,150)(0,0)
\put(30,0){\epsfig{file=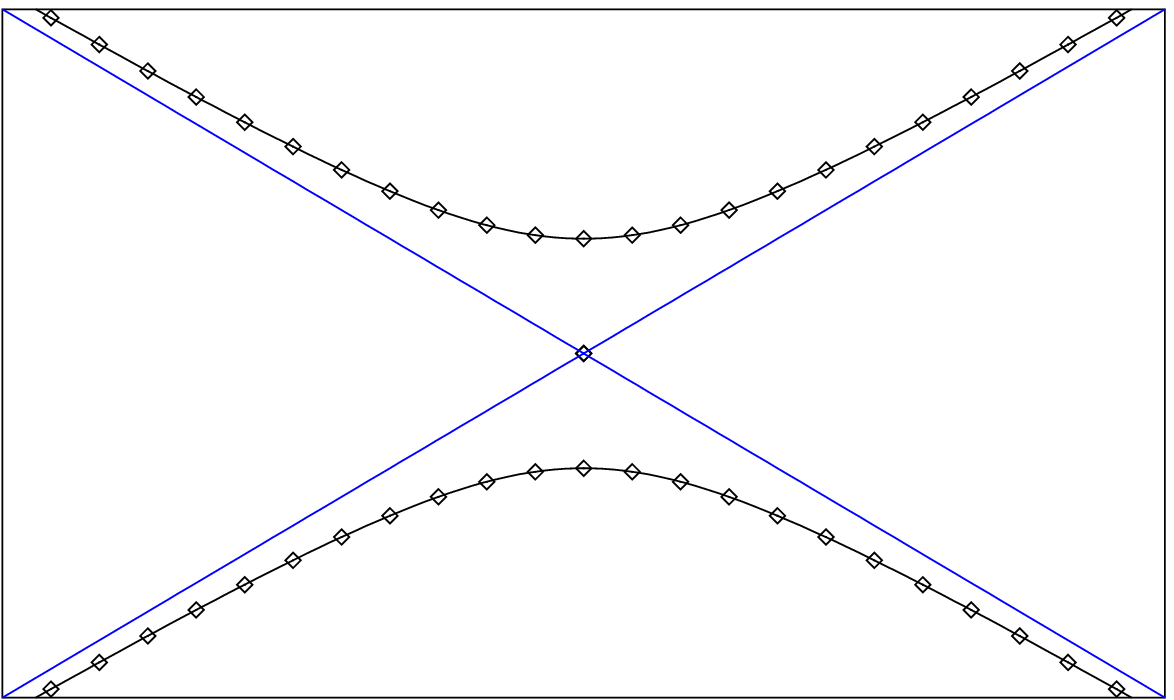,height=5cm}}
\end{picture}
\caption{The point-like branes.
\label{d-1}}}

It is possible to write down the boundary states of these D-branes, at
least at large $k$. The one point functions of these branes are
obtained by evaluating the (properly normalised) closed string primary
vertex operators at the locations of the D-branes. The boundary states
are then superpositions of the Ishibashi states corresponding the the
primaries weighted by the one-point functions.

\subsection{D0-branes}

These are obtained by considering the twined conjugacy classes in the
axially gauged WZW, which are characterised by the class invariant
$Tr(\sigma_1 g)=2 \kappa$.

Thus, projection to the coset gives a connected submanifold, whose
equation in global co-ordinates is $(u-v)=2 \kappa$. Here $\kappa$ can
be any real number. Since the equation describing the world-volume
involves only $(u,v)$ co-ordinates, we can use global co-ordinates to
discuss these branes everywhere in the $(u,v)$-plane, excepting at the
singularities $uv=1$ (analogous to the argument in \cite{Wittenbh}, we
may expect that at the singularity the CFT is well-defined, but the
target space interpretation as a D0-brane fails).

When this is projected down to the coset, we have one relation between
the two co-ords of the coset theory, thus defining a curve. For any
value of $\kappa$ this is a straight line in the $(u,v)$-plane at $45^o$ to
the u(v)-axis. Note that it passes {\it through} the horizons at
$uv=0$.
We can translate these by $l_0={\rm diag}(e^{t_0},e^{-t_0})\,\in H$,
under which the equation $(u-v)=2\kappa$ becomes 
\be u\, e^{t_0} -v\,e^{-t_0} = 2\kappa \ee 
giving us a two parameter family of D0-branes.

\FIGURE{\begin{picture}(300,150)(0,0)
\put(50,0){\epsfig{file=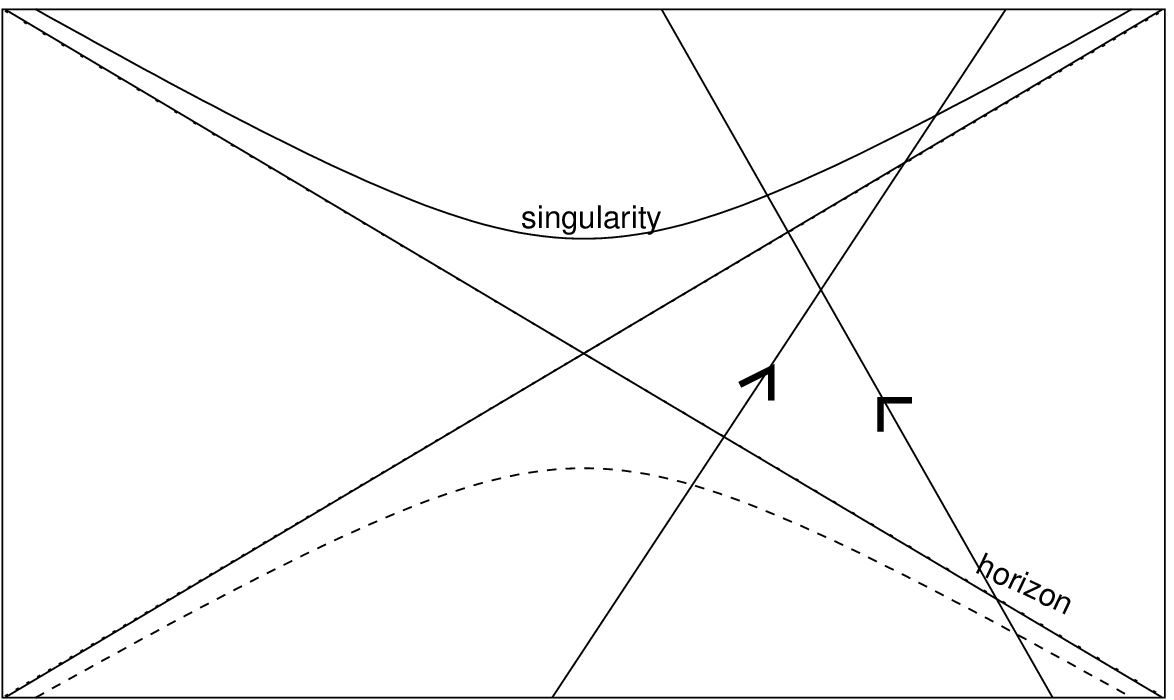,height=5cm}}
\end{picture}
\caption{The D0-brane trajectories.}
\label{D0}}

One can argue for the existence of these branes using the Born-Infeld
action also \cite{MMS}. The effective metric seen by the D0-branes is
\[\frac{ds^2}{g^2 _s}=du\,dv\]
which is flat. Hence, the D0-branes being point particles will move on
geodesics of this metric which are straight lines. It is quite
remarkable that the effective metric is that of {\em flat space} while
the actual background has a curvature singularity even. This tempts
one to speculate that the D0-branes actually pass over to the ``other
side'' of the singularity. 

We can study the minisuperspace spectrum of the strings on the world
volume by using the {\it open string} Laplacian 
\bea 
\Delta_{open} \Psi&=& -\frac{1}{e^{-\Phi}\sqrt{g}}\del_a
\,e^{-\Phi}\sqrt{g} g^{ab}\del_b =\lambda\Psi\label{oplap}\\\nonumber
&=&-k((1-y^2)\frac{d^2}{dy^2}-2y\frac{d}{dy})\Psi \eea

Here $ y=\frac{v+\kappa}{\sqrt{1+\kappa^2}}$ and the singularity
corresponds to $y=\pm 1$ while the horizon is at $y=0,
\frac{\kappa}{\sqrt{\kappa^2+1}}$ (in the equation, $k$ is the level
of the CFT). Also note that $|y|<1$ is equivalent to $uv<1$. 

We recognise the above equation as the Legendre equation. This
equation has two linearly independent solutions $P_\nu(y)$ and
$Q_\nu(y)$, where $\nu$ is a complex number and then
$\lambda=\nu(\nu+1)$. The extended black hole geometry however
corresponds to $y\in [-\infty,\infty]$. (Note that $P_\nu \equiv
P_{-1-\nu}$. We do not have an interpretation of this identity in
physical terms relevant to the D-brane.)

In any case, if $\nu$ is not an integer the $P_\nu$ are all singular
at $y=-1$. This latter point corresponds to the black-hole singularity
(the white hole is the singularity in the first-quadrant).  The
$Q_\nu$ have branch point singularities at $\pm 1$ and $\infty$ if
$\nu$ is not an integer. From the properties of the Legendre
functions, one could guess that that in the $uv<1$ region, the natural
modes are the $P_\nu$ and the $Q_\nu$ are associated with the $uv>1$
region. 

Which modes are physical depends on the boundary conditions
we impose on the wavefunctions. One way  to proceed is by comparison
with the Pf-theory. If we demand regularity at $y=\pm 1$,
then this forces $\nu$ to be an integer. 

Since these branes are time dependent, a natural question to ask is
the meaning of stability and spectrum. The point of the preceding
analysis is that since the effective metric seen by the D0's is flat
and time independent, there is some meaning to the notion of an open
string spectrum. However, the notion of stability is somewhat
unclear. One possible way to characterise these time dependent branes
is that ``nearby'' trajectories stay close (i.e small changes in the
parameters do not lead to divergent effects) \footnote{I am very
  grateful to Ashoke Sen for many illuminating discussions on these points}.


This brane can also be understood in the usual $r,t$-co-ordinates. In
terms of the $(r,t)$-co-ordinates which cover the region in front of
the horizon, the equation defining the brane becomes
\[\sinh{r}\cosh t = 2\kappa \]
Thus as $t\rightarrow \pm \infty$, $r\rightarrow 0$ and as
$t\rightarrow 0$, $r\rightarrow r_{max}=\sinh^{-1} \kappa$. This means
that from the point of view of an asymptotic observer (whose time
co-ordinate is $t$) this brane exited the horizon at $r=0$ infinitely
far back in the past, attained a maximum distance $r_{max}$, and falls
back into the horizon at $r=0$ in the future. She never sees the brane
actually coming {\it out} from the past horizon or {\it crossing} into
the future horizon, as is usual in black hole geometries.

Since this is a D0-brane, we do not have to worry about $B_{NS}$ or
$F$ fields on the world-volume. However, in analogy with \cite{Ashoke}
these branes could carry other conserved charges which could be
calculated by the method described in that paper, once the boundary
state corresponding to these branes is known.


Note that these branes are time dependent, in the sense that the
rolling tachyon is time-dependent. It will be very interesting to
compare the two situations (i.e the rolling tachyon in the Liouville
theory and this) with regard to their dynamics. 

\subsection{D1-branes \label{D1}}

These space-filling branes are obtained by the projections of several
of the regular conjugacy classes to the coset. Many of these
worldvolumes are rather unusual in the sense that they extend behind
the horizon and even behind the singularity. Another surprising
feature is the presence of a boundary even though the bulk geometry is
non-compact. These branes have a world-volume field strength $F$,
turned on for stability.  In 1+1-dimension, a gauge field has no
propogating degrees of freedom. The $F$-field should probably be
thought of as giving rise to a conserved charge which labels the
branes.

In this case, in order to analyse stability of brane, one can study
the Born-Infeld action of the brane with world-volume $F_{uv}$ field
present (the question is whether there is a solution to the BI-
equations of motion of this brane with this F-field which would
(perhaps) stabilise the brane \cite{BDS}).
\bea
S_{DBI}&=& \int e^{-\Phi} \sqrt{-det(g + F)}\\\nonumber
&=&\int e^{-\Phi} \sqrt{\frac{1}{(1-uv)^2}-F_{uv} ^2}
\eea
The equation for F gives
\be
F_{uv} ^2 = \frac{-det(g)\, f^2}{f^2+e^{-2\Phi}}=
\frac{1}{(1-uv)^2}\frac{f^2}{(1+f^2-uv)}\label{F}
\ee
In this equation, $g$ and $\Phi$ are the (pullbacks) of the closed
string metric and dilaton (and we use static gauge in u,v).  It is
easily seen that for both values of $F$, we do not have an imaginary
BI-action i.e the electrical $F$ remains subcritical in the entire
world-volume (in the region $(1-uv)<0$).  Further, note that $F$ blows
up at the singularity $uv=1$ and at $uv=1+f^2$ and, for $uv>1+f^2$, $F$
becomes imaginary. We can interpret this to mean that the D1-branes in
regions I-IV terminate at the singularity, while those in the ``dual''
regions V,VI are bounded by $uv=1$ and $uv=1+f^2$. A precise statement
will however require the construction of the boundary states from
which the boundaries may be inferred.

The $f$ in the equation \ref{F} is proportional to the conjugacy class
trace $\kappa$. One can actually read off the gauge field $F$ from the
boundary terms in the gauged WZW action of the brane as in
\cite{Walton} and obtain the relation between $f$ and $\kappa$.

A possible objection to this procedure is that close to the
singularity curvature effects could become large, and invalidate the
use of the Born-Infeld action.  The leading curvature corrections to
the Born-Infeld action is discussed in \cite{Johnson} (we
refer to the book and the references therein for further
discussion). These take the form 
\be S_{DBI}=\int
e^{-\Phi}\sqrt{-det(\hat g + F)} (1-{\mathcal R}_{abcd}{\mathcal
R}^{abcd}+2 \hat{\mathcal R}_{ab} \hat{\mathcal R}^{ab}) 
\ee
A calculation reveals that both for the Pf-theory and the black hole,
the two curvature terms cancel. Thus at least to this order in
$\alpha'$ the curvature effects do not play a role in this
discussion.

In this section, we shall outline the various regular conjugacy classes
of \sl2 and their projections to the coset which lead to the
space-filling brane world-volumes. 
We shall work with the $(u,v)$ co-ordinates throughout, which is
suitable for describing the branes in the $1-uv>0$ region.

\begin{enumerate}
\item Tr(g)=2\label{tr=2}
\begin{enumerate}
\item $g_0==\left(\begin{array}{cc} 1 & 1\\ 0 & 1 \end{array}\right) $

By explicitly conjugating $g_0$ by the element
$h=\left(\begin{array}{cc} a & u\\-v&b \end{array} \right) \in$ \sl2,
we get the following expression for points in the conjugacy class.
\be p=h\, g_0\, h^{-1}=\left(\begin{array}{cc} a' v'+1 & (a')^2\\
-(v')^2 & 1-a'v'
\end{array}\right)   
\ee 
Thus, from the point of view of the coset, we have
\bea
u= (a')^2\\\nonumber
v=(v')^2\eea
Hence this ``brane'' fills out all of the $(u\geq 0,v\geq 0)$-region of the
coset (because a' and v' range over all of the real line in \sl2). 

\item $g_0=\left(\begin{array}{cc} 1 & 0\\ 1 & 1
\end{array}\right)$ 
In this case, the resultant brane covers the region $(u\leq
0,v\leq 0)$.

\item $g_0=\left(\begin{array}{cc} 1 & -1\\ 0 & 1 \end{array}\right) $
In a similar manner, the points of the conjugacy class are
\be
p=h\, g_0\, h^{-1}=\left(\begin{array}{cc} -a' v'+1  & -(a')^2\\ 
(v')^2 & 1+a'v'
\end{array}\right)   
\ee 
Thus, from the point of view of the coset, we have
\bea
u= -(a')^2\\\nonumber
v=-(v')^2\eea
i.e $u,v \leq 0$.
\item $g_0=\left(\begin{array}{cc} 1 & 0\\ -1 & 1 \end{array}\right) $
In this last case we get the region $u,v\geq 0$.  

We also have conjugacy classes corresponding to the negatives of the
above $g_0$ which project to similar regions.

\end{enumerate}

\item $|Tr(g)|>2$

In \sl2, this c.c forms one connected hyperboloid.

\FIGURE{\begin{picture}(300,150)(0,0)
\put(50,0){\epsfig{file=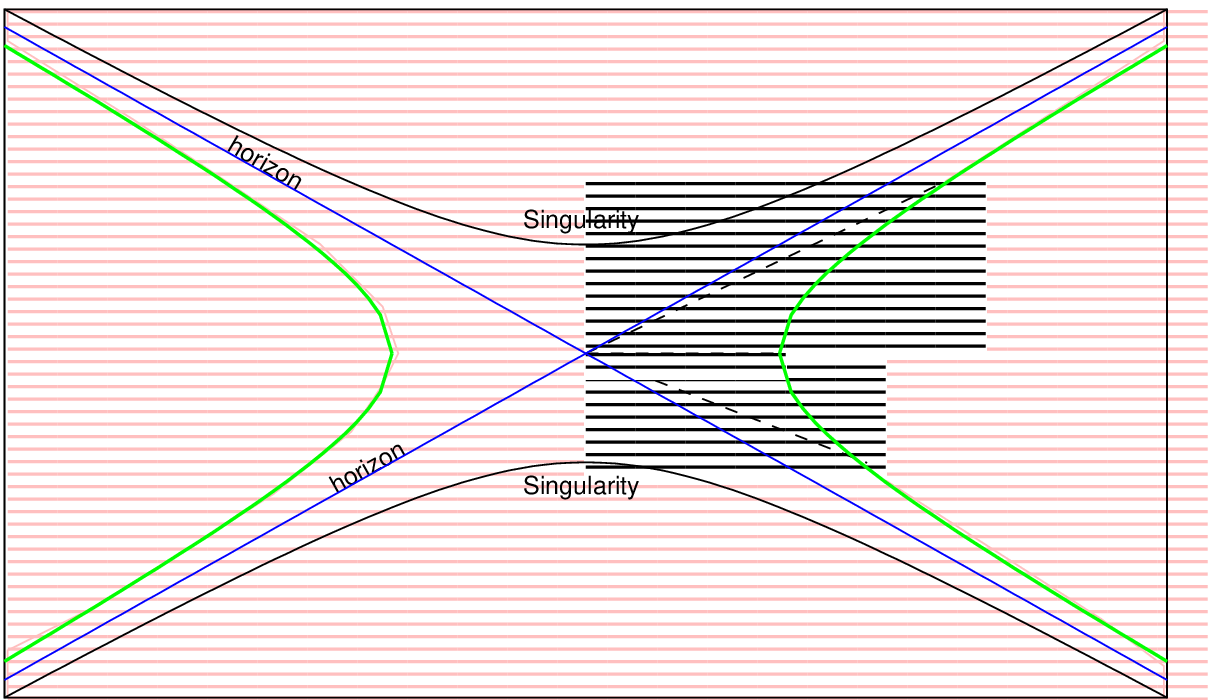,height=5cm}}
\end{picture}
\caption{In-falling D-strings}
\label{D-str}}

Let $g=\left(\begin{array}{cc} x+y
& u\\ -v & x-y \end{array}\right)$, and we are fixing the trace
$2|x|>2$. The determinant condition is \be uv=(1-x^2)+y^2 \ee Thus, as
we vary $y$, this brane covers the region $uv\geq (1-x^2)$. This brane
extends into the physical region $uv<0$, but also covers all of the
region behind the singularity.

\item $|Tr(g)| < 2$

In this case, from the above determinant condition, we see that
$(1-x^2)>0$, which means that the brane covers the region
$uv\geq(1-x^2)>0$ i.e is entirely behind the horizon. 

\FIGURE{\begin{picture}(300,150)(0,0)
\put(50,0){\epsfig{file=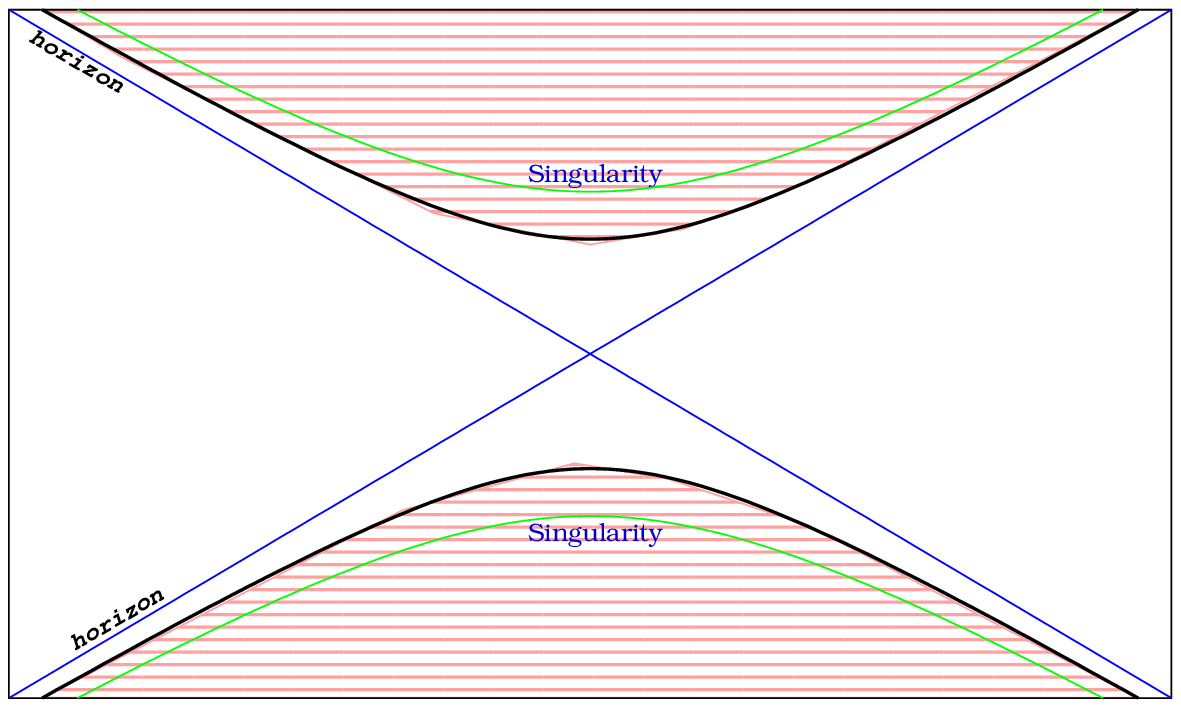,height=5cm}}
\end{picture}
\caption{D2-branes behind the horizon. }}

In the last two cases, we get the same world-volumes for either sign
of the trace $x$. 

\end{enumerate}

Thus we have a family of D1-branes labelled by a single parameter
$\kappa$ corresponding to the trace of $g_0$. The brane world-volumes
are bounded by the singularity at one end on account of the $F$-field
blowing up (it may be noted that $F$ blows up by virtue of its
dependence on the metric). At the other end they are bounded by the
hyperbola $uv\geq 1-\kappa^2$ in regions I-IV, and by $uv=1+f^2$ in
regions V,VI. 

The remarkable thing about these branes is that all the world-volumes
have definite boundaries. A similar situation is seen to occur in the
(euclidean) parafermion theory. In the asymptotically flat region
I,II, since the world-volume gauge field remains finite, it is puzzling 
that the world-volume has a boundary. 

The D1-branes which extend into region I of the black hole geometry
represent world-volumes of D-strings which are emitted by the white
hole and fall back into the black hole. From the point of view of an
asymptotic observer, these D-strings stretch out from the horizon
$\rho=0$ in the far past, extend to a maximum length
$\rho_{max}=1-\kappa^2$ at $t=0$ and then collapse back to the horizon
(in the figure Fig. \ref{D-str}, the dashed lines represent the
D-strings) . Thus these are time dependent and physically observable
to an asymptotic observer.


As in the case of D0-branes, we can study the spectrum of small
fluctuations by performing a minisuperspace analysis. To do this, we
first determine the {\em open} string metric, and coupling. These are
given by the formulae \cite{SW}
\bea
G_{ab}=g_{ab}-(F\,g\,F^{-1})_{ab}\\
G_s=g_s\sqrt{\frac{-det\, G}{-det\,(g+F)}}
\eea
which gives
\bea
G_{ab}&=& \frac{-1}{(1+f^2-uv)}
\left(\begin{array}{cc} 0 & 1\\1&0 \end{array} \right) \\
G_s &=& \frac{g_s}{\sqrt{1+f^2}}\frac{1}{\sqrt{1+f^2-uv}}
\eea
In this case, it is simpler to rescale $u',v'=u,v\sqrt{1+f^2}$ under
which the open string metric and dilaton take the same form as the
bulk values.

As is to be expected, it is seen that when $F_{uv}$ is written in the
{\em open} string co-ordinates, it is (covariantly) constant on the
world-volume consistent with the fact that it is non-dynamical. 

Using these, and the definition of the open string Laplacian given in
\ref{oplap} we can determine the fluctuation modes via the eigenvalue
problem 
\be \Delta_o\Psi(u,v)=\lambda\Psi(u,v) 
\ee 
It proves to be convenient to choose the co-ordinates $y=uv$ and $t$
defined by $-\frac{u}{v}=e^{2t}$, in terms of which the above equation
becomes simple. We shall separate variables by assuming
$\Psi(y,t)=(-y)^{\frac{\pm i\omega}{2}} \,e^{-i\frac{\omega t}{2}}
\Phi(y)$, we obtain the equation governing $\Phi$ as the
hypergeometric equation 
\be
4y(1-y)\Phi''-(\gamma-(\alpha+\beta)y)\Phi'+\alpha\beta\Phi=0 
\ee 
Here $\alpha+\beta=i\omega+1/2$, $\alpha\beta=(\lambda\pm i\omega)/4$
and $\gamma=(i\omega +1)$.  To proceed further and determine
$\lambda$, we need to impose boundary conditions. It is unclear what
are reasonable boundary conditions we should require. In analogy with
the Pf-theory \cite{MMS}, we could require vanishing of the modes at
the singularity $y=1$, or vanishing of the radial derivative at
$y=1$. The latter condition (upon naive analytic continuation) becomes
a Dirichlet condition since $\rho$ is a time-like co-ordinate in this
region.


\section{Comparison with the Pf-theory and Euclidean black hole \label{comp}}

As we remarked in the introduction, all the branes in the Euclidean
black hole geometry constructed in \cite{SR,Foto} have their counterparts
in the Lorentzian background. However, the Lorentzian case has several
BCFT's that do not bear a Euclidean continuation. One can perform an
analysis of the gauged WZW model that leads to the Euclidean hole (or
the Pf-theory) along the lines of this work (\cite{KPY2}) and check
this correspondence.

The $0<uv<1$ region can be analytically continued to the Parafermion
theory, whose branes are discussed in \cite{MMS}. As we discuss below,
all the branes in that theory can be analytically continued subject to
a few modifications.

One point to be noted is that in our work which is at the classical
level, we have not investigated questions about quantization
conditions on the branes. In both the Euclidean theories, the branes
are quantized - a precise comparison of the parameters labelling the
branes will depend on a study of the isometry of the black hole
geometry.

\subsection{D-instantons}
The allowed D(-1)-branes in the Lorentzian black hole are at the
horizon $u=v=0$ and a one parameter family all located on the
singularity $uv=1$.

The $u=v=0$ brane is the continuation of the single D0-brane in
\cite{SR} which sits at the tip of the cigar. The ones on the
singularity are not seen in the cigar geometry. 

The branes on the singularity are the continuations
of the A-branes of \cite{MMS} which are all located on the boundary of
the Pf-target space (which is also a curvature singularity). In that
case, there is a discrete family because of quantization effects (the
rotation isometry of the cigar is anomalous and a $Z_k$ subgroup is
preserved in the quantum theory).

The single brane at $u=v=0$ corresponds to the single B-brane in the
Pf-theory which is located at the centre of the disk.

\subsection{D0-branes}
The embedding equation of the D0-brane world volumes when analytically
continued to the Euclidean case maps into corresponding branes in the
Euclidean geometry \cite{SR}
\[\sinh{r}\cos{\theta}=2\kappa\]
In the Euclidean case, we have two parameter family of these branes. For
reasons similar to the Pf-theory (i.e the rotation isometry is
anomalous) one of the labels is an integer. In the Lorentzian case, we
have two real parameters. This family of branes, in both cases, is in
one-to-one correspondence with the primaries of these theories, in
accordance with the Cardy-correspondence.

Similarly, the embedding equation can be continued to the Pf-theory
also. In this case, we need to either use co-ordinates appropriate to
the region $0\leq uv \leq 1$ (in \sl2 charts).
These D0-branes then map into the D1-branes of \cite{MMS}. 

\subsection{D1-branes}

In the Euclidean black hole case, we have a family of branes labelled
by an integer, which cover the entire cigar as in \cite{SR}
and also another set as described recently in \cite{Foto} which cover
a region near the tip of the cigar. 

These branes are the analytic continuations of the D1-branes of
section \ref{D1}. The latter set corresponds to those c.c with trace
$>2$. However, in the Lorentzian theory, we do not have any D-strings
whose world-volume covers the whole of the $uv<0$ region. It is
tempting to relate the former set of branes in the euclidean theory to
those D-strings with trace $\leq 2$ (If one uses the description of
the {\em Euclidean} black hole as a gauged WZW model, then these
space-filling branes are indeed obtained by projection of those \sl2
c.c with $Tr(g)<2$).

In the Pf-theory, there is a one (discrete) parameter family of
B-branes, which are concentric discs. These are the analytic
continuation of the D-strings.  However, one difficulty is that the
B-branes of \cite{MMS} do not all reach the boundary of the disc while
the ones in the black hole cover the entire $0\leq uv
\leq 1$ region under a naive analytic continuation. However, in both
cases the branes terminate when the $F$-field on the world-volume
blows up. 





\section{Duality relations between the various branes}

The dual geometry in each (i.e Lorentzian and Euclidean) is obtained
by gauging the {\it vectorial} $U(1)$ as opposed to the axial
$U(1)$. In global co-ordinates, as we have discussed briefly in
section \ref{bh}, this gauging implies that the geometry is now
described by the $(a,b)$ co-ordinates. Again $ab=0$ is the horizon
and $ab=1$ is the singularity. 


In the Lorentzian case, if we restrict ourselves to the \sl2
co-ordinate chart relevant to the asymptotically flat region of the
black hole, the dual geometry is the region behind the singularity of
the original black hole.

Since the target space is the same in the dual description, one could
ask how are the D-branes obtained in the vectorial gauging. For this
purpose, it then suffices to examine the various
conjugacy classes in \sl2 and see how they project to the $(a,b)$
plane. The two projections, onto the $(u,v)$ and $(a,b)$ are then to
be understood as being dual to each other. However, under duality the
A-type and the B-type branes are exchanged i.e the branes obtained from 
$Tr(g)=k$ are dual to those from $Tr(\omega g)=k$. This is because the
automorphism $\omega$ takes the $H$-subgroup to its inverse (and the
A-type branes have been defined to be those that are {\em not}
invariant under the target space isometry). 

This however is the same duality relations one obtains in CFT
terms i.e twined conjugacy classes are T-duals (B-branes) of the
regular conjugacy classes (A-branes) (the brane labels are as
appropriate for the vectorially gauged situation).


For instance, the D0-branes in regions I and II are obtained from the
twined conjugacy class of \sl2. Under T-duality, these regions are
mapped to V and VI and the twined c.c become regular conjugacy
classes. These latter are defined by $Tr(g)=a+b=2\kappa$ and are
invariant under the (vectorial) symmetry being gauged. Note that for
various values of the trace, the regular conjugacy classes have fairly
non-trivial geometries in \sl2. But, their {\em projection} to the
coset always yields straight lines in the $(a,b)$ co-ordinates as is
clear from the equation $a+b=2\kappa$. and hence we again obtain
D0-branes. And as before, we can translate these thus giving us a two
parameter family.

There is a special case however; the identity conjugacy class sits at
$a=b=1$ giving point-like D-branes (upon including the translations
we get a one parameter family) as we have already discussed.

In the $(a,b)$ co-ordinates the twined conjugacy classes of \sl2 give
rise to the D1-branes. In this case the determinant condition gives
$ab+v(v+2\kappa)=1$, which implies $ab\leq 1+\kappa^2$. Thus it covers
a region behind the singularity from the point of view of the original
geometry. In the original description, we reasoned that the D-strings
in regions V,VI had to terminate at $uv=1+f^2$ because beyond
$uv>1+f^2$ there was no solution to the B-I equations for $F$. It is
very interesting that the blowing up of $F$ is ``T-dual'' to a $F$
finite situation (this also suggests that $f=\kappa$).

The regions III and IV of the black hole geometry are ``self-dual''
(in the sense the Parafermion theory is self-dual)

These geometrical statements can be made precise by using the
Lagrangian formulation of the brane CFT and following the method of
Buscher \cite{buscher} (for an analysis in the case of the
$SU(2)/U(1)$ see \cite{Walton}).

\section{Discussion}

To summarise, we have performed a detailed analysis of the various
boundary conditions that preserve some part of the current algebra
symmetry of the Lorentzian black hole CFT. 
We have found three kinds of D-branes in this background:
D(-1) branes, D0-branes, and D-strings, several of these occurring in
families. The D0's and D1 branes, from the point of view of an
observer in the asymptotic (flat) region of the black hole appear to
come out of the horizon to a maximum radial distance and then fall
back to the horizon. In global co-ordinates these are emitted by the
white hole and absorbed by the black hole.

We then performed a mini-superspace analysis of the world-volume
theories, and found that the question of the spectrum is not resolved
simply. This is because the allowed open string modes depends on what
boundary conditions we impose at the singularity (for instance). 

We then compared our results with the branes in the Euclidean black
hole and the Pf-theory, and showed that all the branes considered in
these Euclidean theories have their counterparts in the (appropriate
regions) of the Lorentzian black hole.

We conclude by enumerating a number of questions that naturally
arise as a result of this study.

\begin{enumerate}



\item Boundary states: The most important question is to
  construct the wavefunctions that describe these branes. This can be
  achieved in many ways, and perhaps by a judicious combination of
  several of the following.

  We could analytically continue the one-point functions of the
  corresponding Euclidean branes. There is a subtlely in this respect
  that the \sl2 representations in the hyperbolic basis (the principal
  continuous series) appear twice in each unitary irrep of \sl2. This
  makes the analytic continuation somewhat non-trivial.
 
  We could follow the procedure adopted in \cite{MMS} together with
  the known one-point functions of branes in \sl2 to derive the
  boundary states. This procedure is fraught with some difficulty
  because of the non-compactness of the \sl2 CFT.
  
Another method is to use the ``shape of branes'' argument
\cite{Fuchs}. The one-point functions must be such that when projected
onto closed string states which have $\delta$-function wavefunctions,
the amplitudes must be supported on the world-volumes of the
branes. Since we know the geometry, we can derive the one-point
functions in the large $k$-limit, when geometry is reliable by using
the expansion for the delta function in terms of the closed string
vertex operators. 

\item The curvature singularity: It will be of great interest to study
  these BCFT's or equivalently the boundary states to understand the
  singularity. The D-branes we have discussed seem to exist on both
  sides of the singularity; in particular the D0-branes do not seem to
  ``see'' the singularity at all. It remains to be seen how this
  semi-classical result gets modified in a more careful quantum
  analysis of the geometry. 
\item What is the nature, if any, of the closed string
  radiation produced by these branes? The D-branes we have
  constructed are explicitly time-dependent. Once the boundary states
  are known, we can study the nature of the radiation emitted (and
  absorbed) by these, and compare with the Liouville case. The
  presence of the black (and white) hole makes such a calculation
  interesting. Note that while in regions
  I and II the metric is time independent, the metric behind the
  horizon (regions III and IV) is time {\em dependent}. 
\item D0-branes: The D0-branes are described by embedding equations
  $\sinh \rho\, \cosh t=k$ which resemble that of the lump solutions
  of\cite{Sen:2002nu}. It will be of some interest to investigate this
  similarity further, especially the homologue of tachyon matter in
  the black hole geometry. This question could be pursued perhaps
  along the lines of \cite{Gaiotto} where the tachyon matter state has
  been related to an array of branes in imaginary time. 
\item Another natural and important question of study would be the extension of
  these results to the supersymmetric versions of these coset
  theories. The black hole constructions of \cite{harvey} and
  \cite{giddings} has this CFT as a building block. In these cases,
  one can hope that these branes are useful to study the black hole
  singularity in a four dimensional context. 
\item The relationship to the c=1 phase space? The relationship between
$c=1$ string theory and 2D black hole is spelt out in Das \cite{Das:1992dw}
and Dhar et al.\cite{Dhar:1992si}. In particular, the former work
argues that the macroscopic loop equations of the Matrix model is a
transform of the Wheeler-DeWitt equation of the black hole
geometry. This is one way of relating these branes to that of the
$c=1$ string theory. 
\item Recent works \cite{Kutasov,Nakayama} have discussed D-branes in
  the NS5-brane background with space-time behaviour which is similar
  to our D-branes.

\end{enumerate}

{\bf Acknowledgements}: I would like to thank Dileep Jatkar and
Debashish Ghoshal for starting me off on this problem and for
discussions during the initial stages of this work. It is a pleasure
to thank Avinash Dhar, Somdatta Bhattacharya, Rajesh Gopakumar, Namit
Mahajan, Gautam Mandal, L. Sriramkumar and especially Ravi Kulkarni
and Maneesh Thakur for discussions. I am specially grateful to Mathew
Headrick for valuable comments on an initial draft of this manuscript
and to Ashoke Sen and Rajesh Gopakumar for many discussions and
valuable comments on drafts of this work.  I would also like to
acknowledge the use of Matthew Headrick's diffgeom Mathematica
notebook and the hospitality of TIFR, Mumbai and CTS, IISc Bangalore
for hospitality at various stages of this work. This work is supported
by the people of India to whom I am deeply indebted.

\appendix
\section{Appendix A: Co-ordinate systems for \sl2}  

Every matrix $g \in$\sl2 with all entries nonzero can be
written as a product \cite{Vilenkin} 
\be 
g=d_1 (-e)^{\epsilon_1} s^{\epsilon_2} p \,d_2 
\ee 
where $d_{1,2}={\rm diag}(e^{\theta_{1,2}},
e^{-\theta_{1,2}})$ and $\,\,\,\theta_{1,2}\in(-\infty,\infty)$, $e$
is the identity matrix, $s$ is the matrix $\left(\begin{array}{cc} 0& 1\\
        1&0\end{array}\right)$, $p$ is one
of the two matrices
\[ p_1=\left(\begin{array}{cc} \cosh\rho & -\sinh\rho\\
        -\sinh\rho&\cosh\rho\end{array}\right) 
\,\,\,\,\,\rho\in[-\infty,\infty)\]
or 
\[p_2=\left(\begin{array}{cc} \cos\rho &\sin\rho\\
  -\sin\rho&\cos\rho\end{array} \right)
\,\,\,\,\,\rho\in[-\frac{\pi}{4},\frac{\pi}{4}]\]
and $\epsilon_{1,2}=0,1$.

In a similar manner, the matrices in \sl2 with at least one zero entry
can be written as a product
\be
g=d\,(-e)^{\epsilon_1}\,s^{\epsilon_2}\,\left(\begin{array}{cc} 1 &
  0\\ x & 1 \end{array} \right) s^{\epsilon_3}
\ee
where $d={\rm diag}(e^\phi,e^{-\phi})$ and $e$ and $s$ are as above. 

The gauge symmetry that leads to the Lorentzian black hole acts as
$\theta_{1,2}\rightarrow \theta_{1,2}+\epsilon$, and the time
co-ordinate $t$ of the black hole geometry is related to the $\theta$
as $t=(\theta_1-\theta_2)$. It is then easy to see how the various
co-ordinate charts project down (upon gauging) to cover different
regions of the black hole coset.

For instance, setting $p=p_1\,\epsilon_{1,2}=0$ gives us \sl2
matrices of the form
\[g=d_1\left(\begin{array}{cc} \cosh\rho & \sinh\rho\\
        \sinh\rho&\cosh\rho\end{array}\right) \,d_2\] Gauging sets
$d_1=d_2$ and projects to the $(u,v)$ co-ordinates.  The matrices
above are then seen to cover the $uv<0$ regions of the black hole
geometry. Note that $p=p_1,\,\epsilon_1=1,\,\epsilon_2=0$ also covers
the same region of the coset. Similarly, in every co-ordinate chart,
multiplication by the identity matrix yields another copy of the same
region in the coset (upon gauging).  Thus the gauged sigma model gives
two copies of the black hole geometry \cite{Wittenbh}.

Thus we obtain the following covering diagram Fig. \ref{bhcover} (in this
diagram we will omit $\epsilon_1$ since as discussed above, we simply
get another copy of the coset if we include $-I$ factors). The
matrices in \sl2 with zero entries cover the horizon lines $uv=0$ of
the coset and are not indicated in the figure
(Fig. \ref{bhcover}). The singularity is the dark (black) line in the
figure, and the region between the horizon and the singularity is
covered by the two charts with $p_2$ type of matrices. 

\FIGURE{\begin{picture}(300,150)(0,0)
\put(50,0){\epsfig{file=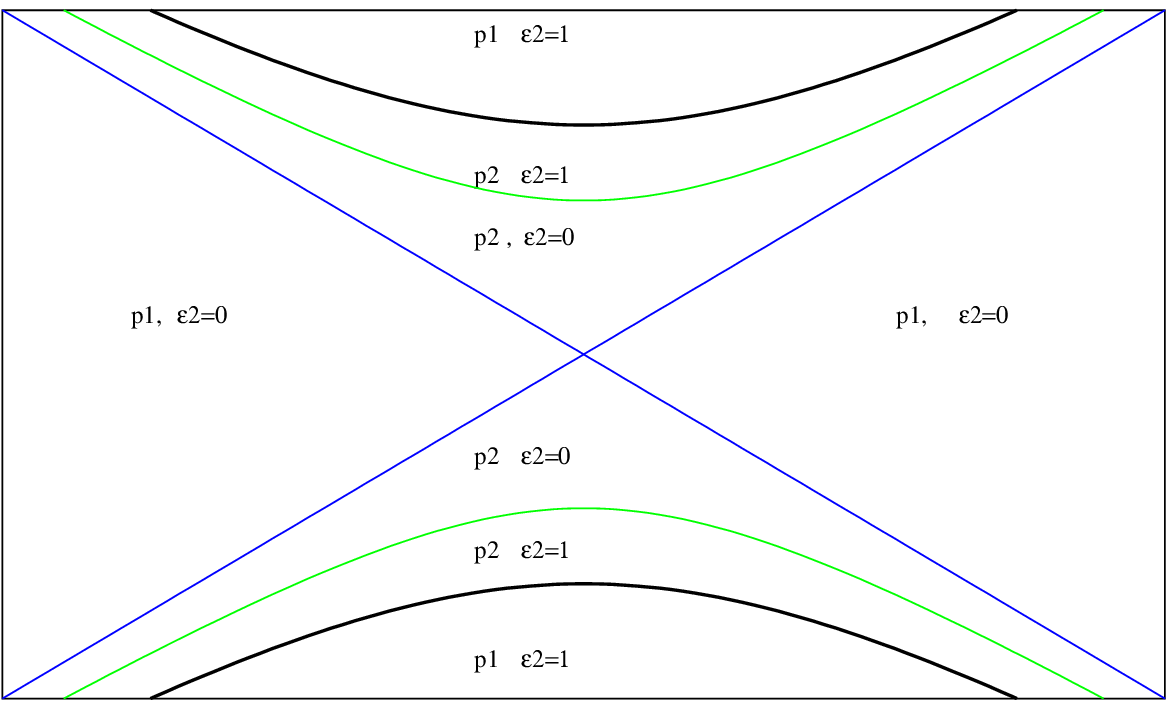,height=5cm}}
\end{picture}
\caption{The black hole in SL(2,R) charts.
\label{bhcover}}}


\section{Appendix C: The BCFT action of the branes}

In this section, we shall briefly discuss the action governing these
boundary conformal field theories. For details, we refer to
\cite{Gawedzki,Elitzur,Walton,KPY2}. Here, we shall consider the vectorial
gauging of a subgroup $H\subset G$ (we can freely switch between the
axial/vector gauging because both lead to the same target space).

The action for a gauged WZW model on a worldsheet without a boundary is 
\bea
S &=& \frac{k_{G}}{4 \pi}\left[\int_{\Sigma}d^2z \,L^{kin} 
+ \int_B \,\omega^{WZ}+S^{\rm gauge} \right]\\\nonumber
&=& \frac{k_{G}}{4 \pi} \left[\int_{\Sigma}d^2z \,Tr(\partial g
  \bar \partial g^{-1} )+\int_B \frac{1}{3}\,Tr(g^{-1} dg)^3\right ]\\\nonumber
&+& \frac{k_{G}}{2 \pi}\int_{\Sigma} d^{2}z \, Tr \left ( \bar A
\partial g g^{-1} - A g^{-1} \bar \partial g +\bar A g A g^{-1}
-A \bar A\right)
\eea
The Wess-Zumino form $\omega^{WZ}$ is integrated over a
3-manifold $B$ whose boundary $\partial b=\Sigma$, the closed-string 
worldsheet. 

In the case of a boundary conformal field theory, since $\Sigma$
itself has a boundary (in our case $\Sigma$ is a disk) there is no $B$
whose boundary is $\partial B=\Sigma$. This action is then modified by
gluing an auxillary disk $D$ to the world-sheet to get a surface
without a boundary, and modifying the action such that the
contributions from the disk $D$ cancel. A further requirement that the
various embeddings the disk in $G$ should give the same contribution
then restricts the allowed boundary values further. This is achieved
as follows.

For the A-type branes, the worldsheet field $g(z,\bar z)$ is extended
to the disk $D$ by requiring that on $D$ and on the boundary $\partial
\Sigma$  g is restricted to the set $g \in C^G _f\, C^H_l$, i.e a product
of conjugacy classes $C^G_f=kfk^{-1}$ of $G$ and $ C^H_l=plp^{-1}$ of
$H$ respectively. 

The additional term has the form
\be
-\frac{k_G}{4\pi}\int_D\,\Omega^{(f,l)}(k,p)=-\frac{k_G}{4\pi}\int_D\,
\left[ \omega^f(k)+Tr(dc_2 c_2 ^{-1} c_1
  ^{-1} dc_1)+\omega^l(p)\right]\label{add}
\ee
Here $c_1= kfk^{-1}, f,k\in G$ and $c_2= plp^{-1}, p,l\in H$ and 
$\omega^g(h)$ is the Recknagel-Schomerus two form
\[ \omega^g(h)=Tr(h^{-1} dh g h^{-1} dh g^{-1})\]

The two form $\Omega^{(f,l)}(k,p)$ in the above integral has the
property that $d\Omega^{(f,l)}(k,p) =w^{WZ}(g)$ when $g$ is restricted
to the set $C^G _f \,C^H_l$; hence on the auxillary disk the two terms
cancel. The boundary conditions on $g$ are that it is restricted to a
product of the conjugacy classes, except that the gauging identifies
field configurations $g\sim h\,g\,h^{-1}$ for $h\in H$. Thus, the
worldsheet boundaries are restricted to the set $C^G _f C^H_l$
projected to the coset $G/H$.

The case of the B-type branes is similar \cite{Walton,KPY2}. In this
case the boundary conditions on the field $g$ restricts the endpoint
to lie in product of { \em twined} conjugacy classes $C^{\alpha,G}
_f\, C^{\alpha,H}_l$ where $\alpha$ is an automorphism of $G$.
Correspondingly, the additional terms \ref{add} that were required to
cancel the contribution of the WZ three form on the auxillary disk
have a different form. In this case the extra pieces of the action are
\be 
-\frac{k_G}{4\pi}\int_D\,\Omega^{(f,l)}(k,p)=-\frac{k_G}{4\pi}\int_D\,
\left[ \omega^f(k)+Tr(dc_2 c_2 ^{-1} c_1^{-1} dc_1)+\omega^l(p)\right]
\label{addb}
\ee

In this case, the two form $\omega$ is given by
\[ \omega^g(h)=Tr(\alpha(h^{-1} dh) g h^{-1} dh g^{-1})\] and 
$c_1= \alpha(k)fk^{-1}, f,k\in G$ and $c_2= \alpha(p)lp^{-1}, p,l\in
H$ As before, on the set $C^{\omega,G} _f\, C^{\omega,H}_l$,
$d\Omega^{(f,l)}(k,p) =w^{WZ}(g)$ and the two terms cancel.


In a similar manner, we can work out the case when the gauged symmetry
is the axial embedding of $H$. As one could perhaps guess, the A- and
B-type branes are interchanged. In our case, $G=SL(2,R)$ and $H=U(1)$,
and both the axial and vectorial gauging give rise to the same target
space geometry.

\end{document}